\documentclass[10pt,conference]{IEEEtran}
\IEEEoverridecommandlockouts
\usepackage{cite}
\usepackage{amsmath,amssymb,amsfonts} 
\usepackage{algorithmic}
\usepackage{graphicx}
\usepackage{textcomp}
\usepackage{xcolor}
\usepackage{comment}
\usepackage{hyperref}
\usepackage{xcolor}
\usepackage{booktabs} 
\usepackage{colortbl}
\usepackage{longtable}
\usepackage{balance}

\def\BibTeX{{\rm B\kern-.05em{\sc i\kern-.025em b}\kern-.08em
    T\kern-.1667em\lower.7ex\hbox{E}\kern-.125emX}}
\begin{document}

\makeatletter 
\newcommand{\linebreakand}{%
  \end{@IEEEauthorhalign}
  \hfill\mbox{}\par
  \mbox{}\hfill\begin{@IEEEauthorhalign}
}
\makeatother 

\title{MONO2REST: Identifying and Exposing Microservices: a Reusable RESTification Approach}

\author{\IEEEauthorblockN{Matthéo Lecrivain}
\IEEEauthorblockA{\textit{Nantes Université, LS2N UMR 6004},\\ F-44000 Nantes, France \\
mattheo.lecrivain@tuta.io}
\and
\IEEEauthorblockN{Hanifa Barry}
\IEEEauthorblockA{\textit{DIRO, Université de Montréal}\\
Montréal (QC), Canada \\
hanifa.barry@umontreal.ca}
\and
\IEEEauthorblockN{Dalila Tamzalit}
\IEEEauthorblockA{\textit{Nantes Université, LS2N UMR 6004},\\ F-44000 Nantes, France \\
Dalila.Tamzalit@univ-nantes.fr}
\and
\linebreakand

\IEEEauthorblockN{Houari Sahraoui}
\IEEEauthorblockA{\textit{DIRO, Université de Montréal}\\
Montréal (QC), Canada \\
sahraouh@iro.umontreal.ca}
}

\maketitle

\begin{abstract}

The microservices architectural style has become the de facto standard for large-scale cloud applications, offering numerous benefits in scalability, maintainability, and deployment flexibility. Many organizations are pursuing the migration of legacy monolithic systems to a microservices architecture. However, this process is challenging, risky, time-intensive, and prone-to-failure while several organizations lack necessary financial resources, time, or expertise to set up this migration process. So, rather than trying to migrate a legacy system where migration is risky or not feasible, we suggest exposing it as a microservice application without without having to migrate it. In this paper, we present a reusable, automated, two-phase approach that combines evolutionary algorithms with machine learning techniques. In the first phase, we identify microservices at the method level using a multi-objective genetic algorithm that considers both structural and semantic dependencies between methods. In the second phase, we generate REST APIs for each identified microservice using a classification algorithm to assign HTTP methods and endpoints. We evaluated our approach with a case study on the Spring PetClinic application, which has both monolithic and microservices implementations that serve as ground truth for comparison. Results demonstrate that our approach successfully aligns identified microservices with those in the reference microservices implementation, highlighting its effectiveness in service identification and API generation.

\end{abstract}


\section{Introduction}
\label{Introduction}
Legacy software systems are applications developed using outdated architectural principles and technologies, rendering them technically obsolete despite their continued functional relevance. Most legacy systems rely on a monolithic architecture, where the application is a single, integrated entity encompassing all functionalities. However, monolithic legacy systems face significant challenges. They often lack scalability, struggle to improve performance, and fail to quickly incorporate new features, making them ill-suited to meet evolving market demands~\cite{mahanta2020translating}. Additionally, the use of deprecated technologies increases the likelihood of bugs, necessitating specialized knowledge of the system and its dependencies. The inherent complexity of the code further hinders upgrades to modern technologies, particularly cloud-native architectures. 

Maintaining such legacy systems becomes increasingly burdensome over time~\cite{lewis2005service}. Addressing these issues requires migrating these systems to newer technologies and modern architectural styles. Among these, microservices architecture (MSA) has emerged as a dominant paradigm in software engineering~\cite{Fowler2014}. Designed for the cloud, MSA enables independent deployment of services, with communication facilitated through APIs~\cite{dragoni2017microservices}. This architecture improves scalability, simplifies maintenance, and enhances agility, as changes can be made to individual services without affecting the entire application~\cite{Fowler2014, asrowardi2020designing}.

Despite its advantages, migrating monolithic legacy systems to microservices is a complex and resource-intensive process that requires specialized expertise, significant time, and substantial investment~\cite{newman2021building, gouigoux2017monolith, gouigoux2019functionalfirst}. Moreover, not all organizations possess the readiness, budget, or skills necessary for such a transition~\cite{gouigoux2021microservice}. For these cases, existing solutions fail to offer alternatives that enable legacy systems to operate as microservices-based applications without full migration.

This paper presents a novel approach that addresses this gap by exposing monolithic legacy systems as microservices-based applications without requiring migration. Our reusable, automated, two-phase framework combines evolutionary algorithms with machine learning techniques. In the first phase, a multi-objective genetic algorithm identifies microservices at the finest granularity—the method level—by analyzing both structural and semantic dependencies. In the second phase, REST APIs are generated for each identified microservice using a classification algorithm to assign HTTP methods and endpoints.
To evaluate our approach, we conducted a case study using an application with both monolithic and microservices implementations. The results show that the identified microservices from the monolithic version align closely with those in the microservices version, and the generated REST APIs effectively expose the services in a consistent and functional manner.

This paper is organized as follows: Section~\ref{BackgroundIssue} provides the background and problem context. Our two-phase approach for microservice identification and REST API generation is detailed in Section~\ref{Approach}, while Section~\ref{Validation} presents the case study of the Spring PetClinic application. Section~\ref{RelatedWorks} reviews related work in this domain. Section~\ref{Discussion} discusses our findings, and Section~\ref{Conclusion} concludes the paper and proposes ideas on future work.

\section{Background and Problem Statement}
\label{BackgroundIssue}
There are many research contributions targeting the detection of microservices in monolithic architectures and migrating them, or preparing them for migration, to this architectural style~\cite{gouigoux2017monolith, newman2021building, balalaie2018microservices, taibi2017processes, de2020remodularization}. Migrating a legacy software system to a microservices architecture involves restructuring it into independent services, each representing a distinct functionality of the original application. These services function as standalone software applications, and their simultaneous operation produces functional results equivalent to traditional architectures~\cite{dragoni2017microservices}. Proposed solutions generally divide the migration process into two distinct phases: \textit{service identification} and \textit{application migration}.

Service identification involves partitioning and decomposing the original application into clusters, which then serve as a foundation for creating services. These clusters must align with business logic and be capable of autonomous operation. This phase is complex and requires a thorough understanding of the monolithic application's internal workings, including the relationships and dependencies among its components. In the application migration phase, the monolithic application is transformed into a microservices architecture based on the identified clusters. Although several studies have explored the challenges of automated approaches to migration~\cite{migration-challenges, SurveyMigration}, to the best of our knowledge, no fully implemented solution currently exists. Automated migration faces several challenges: \textit{(i)} converting a monolithic system into a microservices-based architecture requires significant rewriting and reorganization of code, which demands an in-depth understanding of the system, \textit{(ii)} the generated services must meet high-quality standards and be compatible with the target environment while preserving core functionality and maintaining optimal performance, \textit{(iii)} dependencies between services, whether through method calls or shared data, must be carefully managed to ensure smooth operation and prevent communication conflicts.

Given these challenges, not all monolithic systems are suitable for migration to a microservices architecture. The process is inherently challenging, risky, time-consuming, and prone to failure~\cite{zhang2024failure}. It requires significant financial resources, time, and expertise that many organizations lack. Furthermore, organizations may struggle to assess their readiness for migration~\cite{gouigoux2021microservice}. Migrating poorly designed monoliths can also negatively affect the resulting microservices architecture by transferring technical debt, design flaws, and dead code into the new system~\cite{mahanta2020translating}. Additionally, migration must preserve the application's integrity, including its data. These difficulties often lead to the persistence of legacy systems, which remain constrained by their monolithic architectures due to a lack of viable alternatives.

In this context, companies face a binary choice: either retain their monolithic applications or fully migrate to a microservices architecture. Currently, no alternative solutions exist that balance the benefits of microservices with the risks and costs of migration. To address this gap, we propose a novel approach that enables monolithic systems to function as microservices-based architectures without requiring structural changes. Wrapping these systems with REST APIs offers several advantages: it allows organizations to evaluate the feasibility of continued use in their current form, assess their readiness and suitability for future migration, act as an intermediate step in the migration process, and identify potential challenges that could hinder a successful transition.

The context of migrating monolithic legacy systems to adopt a microservices architectural style involves several dimensions. The following assumptions delimit the scope of our approach and position the addressed issues. We focus on object-oriented monolithic software developed in Java and built using Maven or Gradle project architectures. This ensures that the project structure consistently follows a standard format, enabling operations and providing a reusable approach for legacy systems within the same ecosystem. Within this context, we address two main issues:
\begin{enumerate} 
\item \textit{MS granularity and semantic:} While service identification has been widely studied, many approaches fail to consider the optimal level of granularity and semantic similarity when creating clusters. Most approaches focus on class-level decomposition, but service identification should target behavior at the method level. Methods from different classes can contribute to the same functionality, and conversely, methods within the same class may belong to different microservices~\cite{abgaz2023decomposition}. Additionally, most studies prioritize structural dependencies in the code while overlooking business semantics, which play a critical role in accurately identifying microservices~\cite{gouigoux2019functionalfirst}. \item 
\textit{Alternative solution to migration:} To the best of our knowledge, no existing work has proposed exposing monolithic software systems on the web by wrapping them with automatically generated REST APIs without requiring migration. 
\end{enumerate}
In this paper, we address these issues by 1) leveraging a multi-objective genetic algorithm to identify microservices based on structural and semantic dependencies at the method level, and 2) exposing the identified microservices through automatically generated REST APIs using machine learning techniques to assign HTTP methods and endpoints.

\section{Approach}
\label{Approach}
This section describes the two stages of our approach. First, we describe our method for identifying microservices in a monolithic application, using a genetic algorithm and machine learning. Then, we present our interface identification approach for exposing the identified services, combining static analysis and several machine-learning algorithms and models.

\subsection{Microservices identification}
\label{ssec:MSIdentification}

\begin{figure*}[h]
    \centering
    \includegraphics[width=.7\textwidth]{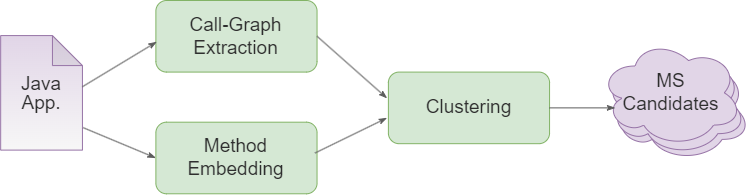}
    \caption{Service identification process.}
    \label{fig:micid}
\end{figure*}

The primary challenge in migrating legacy monolithic systems lies in accurately identifying microservices with appropriate granularity, ensuring they reflect both the structural and semantic aspects of the original system~\cite{gouigoux2017monolith, newman2021building, balalaie2018microservices, taibi2017processes}. To address this, we propose a microservice identification approach based on a multi-objective genetic algorithm. As illustrated in Fig.~\ref{fig:micid}, the identification process consists of three key components: call-graph extraction, which captures the structural dependencies within the monolithic system; method embedding, which represents semantic relationships between methods; and clustering, which groups methods into microservice candidates.

\subsubsection{Call-Graph Extraction}

The call graph is extracted from the JAR file of the monolithic application. This process begins by collecting all method calls using the static analyzer \textit{java-callgraph}\footnote{\url{https://github.com/gousiosg/java-callgraph}}, which outputs a list of sources and destinations for all class and method calls (see an example in Fig.~\ref{fig:micid:extraction}).

\begin{figure}[h]
    \centering
    \includegraphics[width=0.5\textwidth]{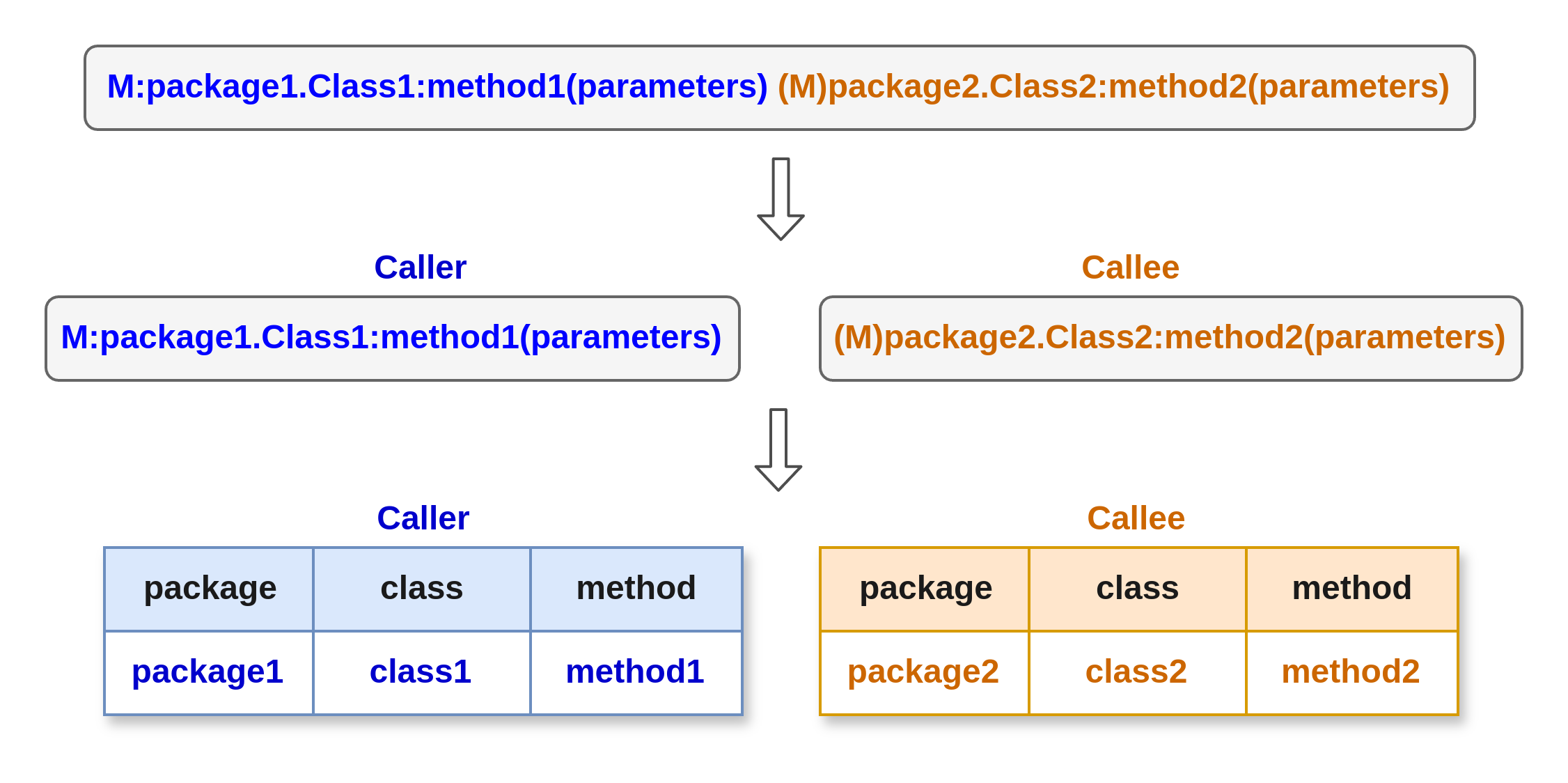}
    \caption{Preparation of data linked to method calls.}
    \label{fig:micid:extraction}
\end{figure}

Next, the list is refined by removing class calls, JDK method calls, and recursive method calls, retaining only the dependencies between the application's methods. This refinement produces a directed call graph. 
The resulting call graph serves as the basis for clustering, as it will be partitioned to identify potential microservices. Additionally, it provides a visual representation of the candidate microservices. Fig.~\ref{fig:micid:dependency_graph} illustrates the call graph derived from the Spring PetClinic application, which will be used in our case study.

\begin{figure}[h]
    \centering
    \includegraphics[width=0.5\textwidth]{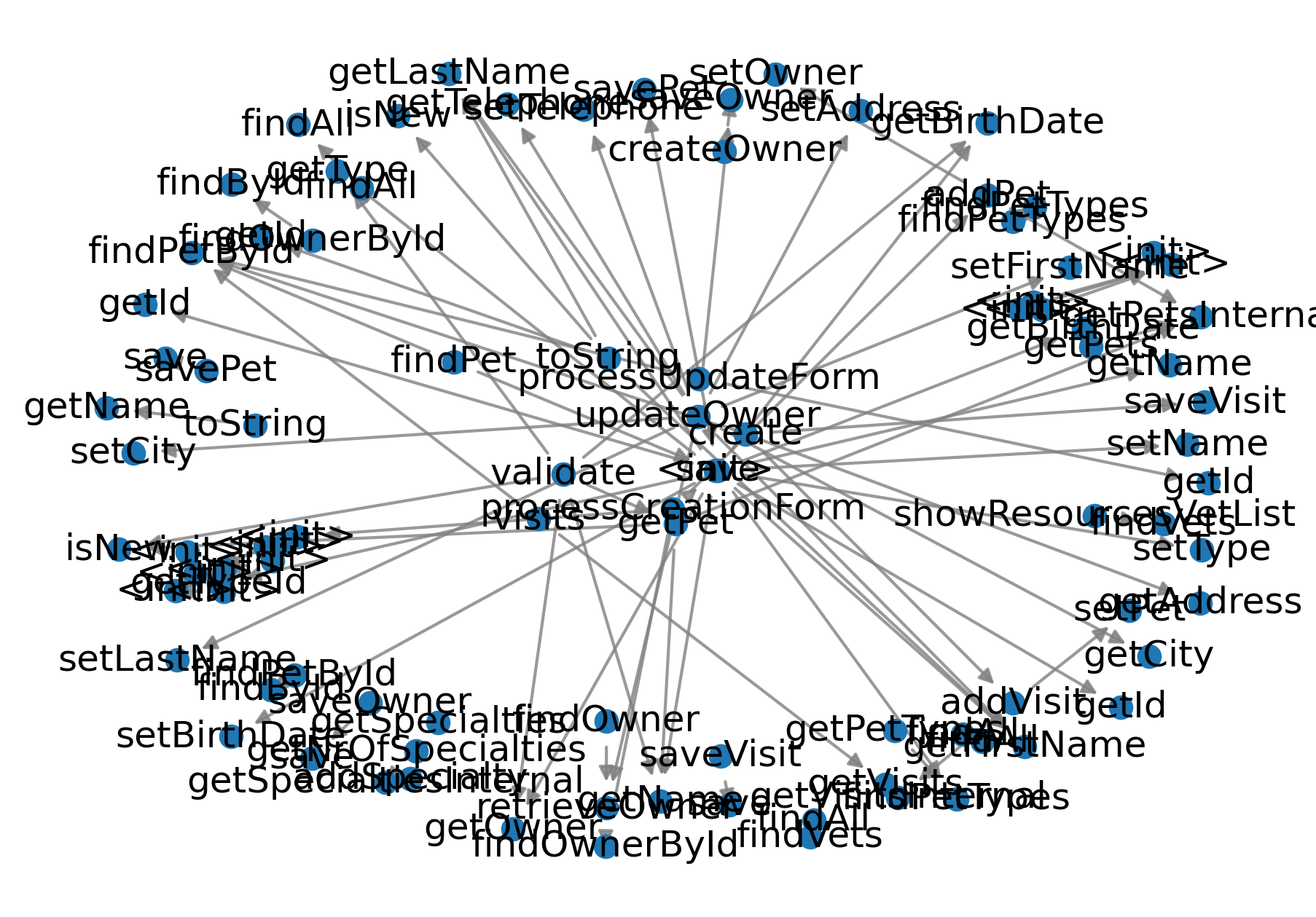}
    \caption{Call graph for the Spring PetClinic application.}
    \label{fig:micid:dependency_graph}
\end{figure}

The structural dependencies within the call graph are used to calculate the \textit{coupling} and \textit{cohesion} between candidate clusters, which represent potential microservices. Coupling measures the dependencies between methods in two different clusters. For example, in Fig.~\ref{fig:micid:clusters}, there is only one dependency between Cluster A and Cluster B, represented by the call from $m_5$ to $m_2$.
Cohesion, on the other hand, captures the internal dependencies between methods within the same cluster. For instance, in Cluster A, there are three calls among its four methods.
The specific calculations for coupling and cohesion, as well as their role in the clustering process, are detailed in the description of the genetic algorithm.

\begin{figure}[h]
    \centering
\includegraphics[width=0.45\textwidth]{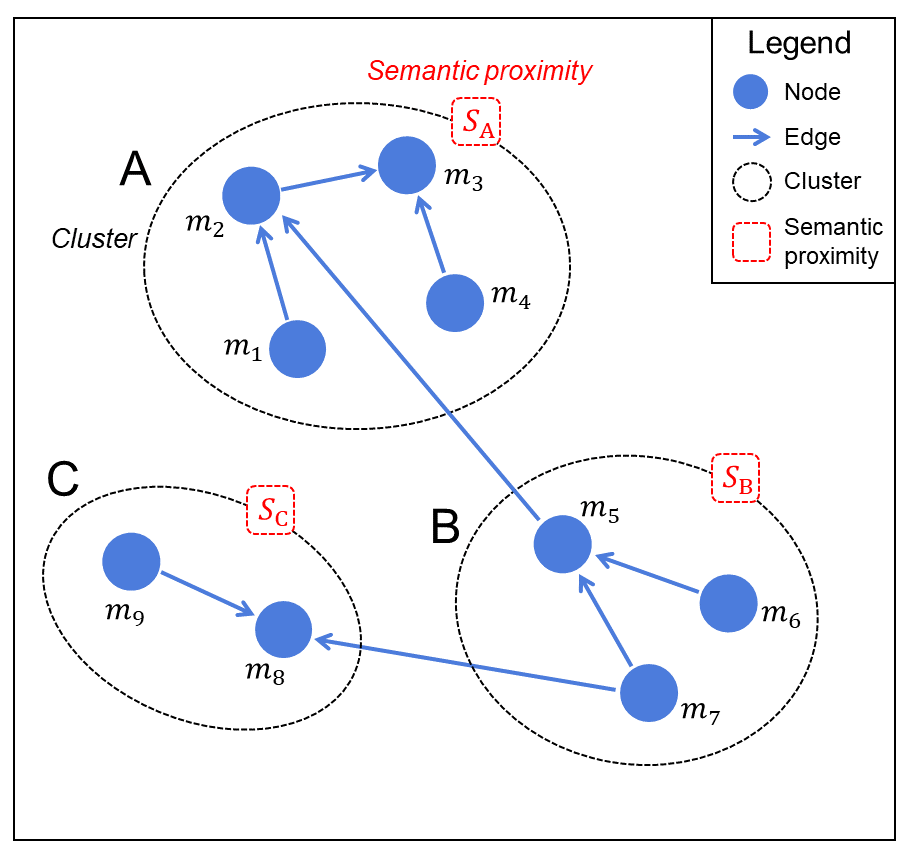}
    \caption{Example of a directed dependency graph organized in clusters.}
    \label{fig:micid:clusters}
\end{figure}

\subsubsection{Method Embedding} 

To capture the semantic relationships between methods, we employ a sentence embedding technique. The objective is to determine whether methods within a cluster, representing a potential microservice, are semantically related. There are two approaches for generating the method embeddings: method names only that consists in extracting terms from method names after tokenization, or full method context with the inclusion of all identifiers present in the method, such as parameter and variable names, in addition to method names.
The list of extracted terms is then processed using an embedding model\footnote{\url{https://huggingface.co/sentence-transformers/bert-base-nli-mean-tokens}} \cite{reimers2019sentencebert, thakur2021augmented}, which generates numerical vectors for each method. The semantic similarity between two methods is determined by the distance between their respective vectors: the smaller the distance, the stronger the semantic relation between the methods.

\subsubsection{Clustering} Before performing the clustering, we first encode the call graph and the similarities between methods in data structures that facilitates the clustering process. The call graph is encoded as an adjacency matrix, where a value of 1 in a cell signifies a dependency between the two corresponding methods. Additionally, a second matrix encodes the semantic similarities between each pair of methods in the application. Each cell in this similarity matrix contains the cosine similarity, ranging from -1 to 1, between the vector representations of the two methods. Precomputing these similarities accelerates the iterative clustering process. Table~\ref{fig:similarity_matrix} presents an excerpt of the similarity matrix from the application Spring Pet Clinic, highlighting a strong semantic similarity (\mbox{$\approx0.932$}) between the methods \texttt{findPetById} and \texttt{findOwnerById}.

            \begin{table}\centering
            \begin{tabular}{l|ccc|}
                    & \textbf{findPetById}   & \textbf{findOwnerById}  & \textbf{setOwner} \\ \midrule
                \textbf{findPetById}  & $\approx 1.000$ & $\approx 0.932$ & $\approx 0.502$ \\
                \textbf{findOwnerById}  & $\approx 0.932$ & $\approx 1.000$ & $\approx 0.597$ \\ 
                \textbf{setOwner}  & $\approx 0.502$ & $\approx 0.597$ & $\approx 1.000$ \\
                \bottomrule
            \end{tabular}
            \caption{Subset of the similarity matrix of the Spring PetClinic application.}
            \label{fig:similarity_matrix}
            \end{table}
            

A clustering solution represents the partitioning of the set of $n$ methods in the application into $k$ subsets, where each subset corresponds to a potential microservice. The size of the search space is determined by the Stirling number of the second kind ${n\brace k}$. Given the exponential growth of this number, exhaustive exploration of the search space is infeasible, making heuristic algorithms essential for effectively navigating and optimizing the solution space.

To perform the clustering that generates microservice candidates, we leverage the latest version of the NSGA multi-objective optimization algorithm, NSGA-III\cite{NSGA-III}, implemented in the pymoo library\cite{pymoo}. 
NSGA-III is designed to handle problems with three or more conflicting objectives effectively.
It starts by the generation of a population of random solutions, in our case a set of different clustering solutions. Then in each iteration, the population of solutions undergoes non-dominated sorting to rank solutions based on Pareto dominance, dividing them into fronts. The algorithm uses reference points in the objective space to guide the selection process, ensuring diversity across the Pareto front. Solutions are then evolved through crossover and mutation to generate offspring, which are combined with the parent population. From this combined set, the next generation is selected based on both rank (non-domination level) and distance to reference points, favoring solutions near underrepresented points. This iterative process continues until a stopping criterion, such as a maximum number of generations, is met. 

\begin{figure*}[h]
    \centering
\includegraphics[width=0.65\textwidth]{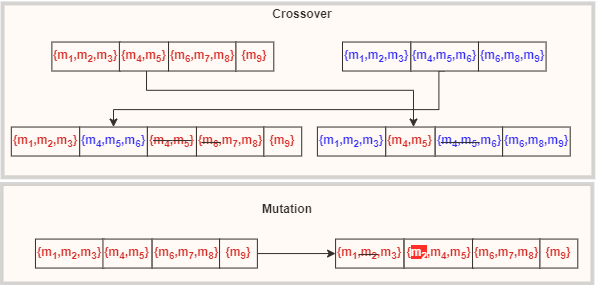}
    \caption{Crossover and mutation operators.}
    \label{fig:micid:operators}
\end{figure*}

As illustrated in Fig.~\ref{fig:micid:operators}, the crossover operator involves injecting a cluster from one parent into the second parent and repairing the clusters to prevent method duplications. For the mutation operator, a method is randomly selected and transferred to another cluster.

To evaluate the quality of a clustering solution and select solutions for the next iterations, we use three objective functions. Let us begin with some definitions:
\begin{itemize} 
\item $V_{cluster}(x)$: Number of nodes (methods) within a cluster $x$. 
\item $E_{internal}(x)$: Number of links (calls) between nodes within a cluster $x$. 
\item $E_{external}(x)$: Number of links connecting nodes in cluster $x$ to external clusters. 
\item $Sem(a,b)$: Semantic similarity between two methods $a$ and $b$. 
\item $N$: Number of clusters within a solution $S$. 
\end{itemize}
The three clustering objectives are defined as follows:

\begin{enumerate} 
\item \textbf{Minimize coupling} between clusters, i.e., the connectivity between methods belonging to different clusters. The coupling value of a cluster $x$ is defined as the proportion of external dependencies relative to the total dependencies of the methods in the cluster. Formally:

\[ Coupling(x) = \frac{E_{external}(x)}{ E_{internal}(x) + E_{external}(x) }  \]
The coupling for a clustering solution $S$ is the sum of the coupling values of its clusters:

\[ Coupling(S) = \sum_{x_i \in S} Coupling(x_i) \]

\item \textbf{Maximize cohesion} within clusters, i.e., the connectivity between nodes of the same cluster. The cohesion value of a cluster $x$ is defined as the number of internal dependencies normalized by the size of $x$. Formally: 

\[ Cohesion(x) = \frac{E_{internal}(x)}{V_{cluster}(x)} \]
If the number of internal dependencies exceeds the size of the cluster, the cohesion is set to $1$. The cohesion of a clustering solution $S$ is calculated as the average cohesion of its clusters:
\[ Cohesion(S) = \frac{1}{N} \sum_{x_i \in S} Cohesion(x_i) \]

\item \textbf{Maximize semantic similarity} between methods within clusters. For each cluster $x$, we use the similarity values $Sim(i,j)$ from the similarity matrix for pairs of methods $(i,j)$ in $x$. The semantic similarity of a cluster $x$ is the average similarity between pairs of methods in $x$:

\[ SemSim (x) = \frac{1}{V_{cluster}} \sum_{i, j \in x \wedge i \neq j} Sim(i, j) \]
The semantic similarity of a clustering solution $S$ is then calculated as the average semantic similarity of its clusters:
\[ SemSim (S) = \frac{1}{N} \sum_{x_i \in S} SemSim(x_i) \]
\end{enumerate}




\subsection{Interfaces identification and assignment}
\label{ssec:InterfacesIdentification}
We now detail the second phase of our approach, which focuses on identifying interfaces to expose the microservices previously identified. The objective is to define a REST API for each cluster, enabling the application to be wrapped instead of fully migrated. This approach offers an alternative to complete application migration by simulating autonomous and independent services while retaining the monolithic structure.
The API identification process combines regular expression-based parsing with machine learning models. Fig. \ref{fig:restify} illustrates the pipeline for interface identification, a process we call "RESTification." This term refers to the creation of "RESTful" services, which are exposed through hierarchical interfaces and communicate using CRUD operations \cite{10.1145/3486608.3486913, Fielding2000ArchitecturalSA}.

\begin{figure*}[h]
    \centering
    \includegraphics[width=0.8\textwidth]{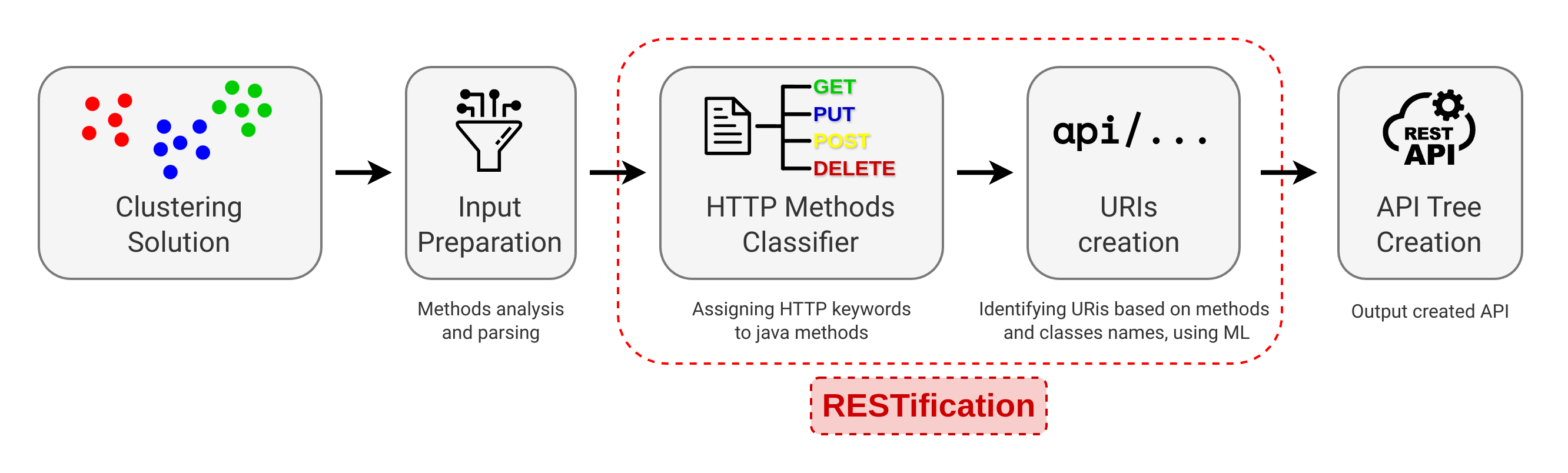}
    \caption{Interface identification process.}
    \label{fig:restify}
\end{figure*}

The first step of the pipeline consists in selecting for each cluster the methods that should be exposed through an API together with their respective contexts.  Fig.~\ref{fig:restify:call_types} shows an example where method $m_1$ is called by $m_2$, which belongs to a different cluster. In this case, $m_1$ must be exposed through the API of cluster $A$, allowing method $m_2$ to call it directly via the API. This subset of methods to expose is therefore defined as the set of methods within a cluster that are invoked by methods external to the cluster.
Once a method is selected, we gather additional information about it. The algorithms used in subsequent steps require this contextual data about the methods to make informed decisions. Specifically, for each method, we extract its return type and the types of its parameters. To achieve this, the application’s JAR file is extracted and analyzed using the \texttt{javap} command, which disassembles \texttt{.class} files and generates a file containing all class and method definitions within the application. Each selected method is then defined as ''\texttt{return\_type method(parameters\_types)}"

\begin{figure}[!h]
    \centering
    \includegraphics[width=0.4\textwidth]{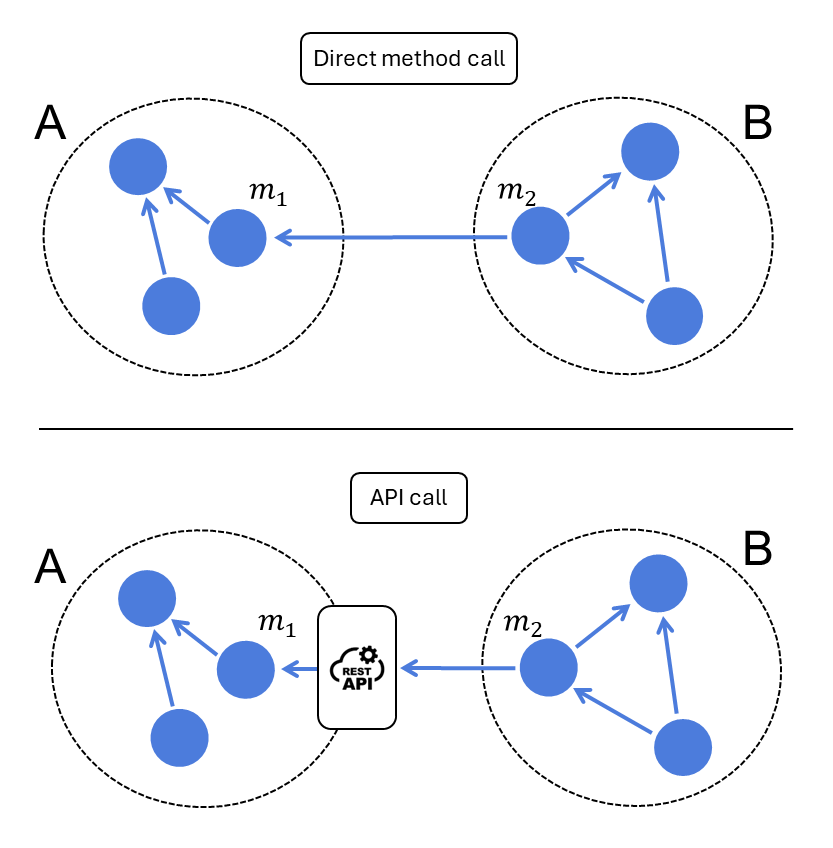}
    \caption{Comparison between a direct method call and a call to an API at cluster level.}
    \label{fig:restify:call_types}
\end{figure}

Interactions with a REST API are conducted through HTTP requests. To perform these requests, it is necessary to determine the API "endpoint," which specifies the address where the request will operate, the type of HTTP request, and the request's content.
This step focuses on determining the type of HTTP request to associate with each method that needs to be exposed. We consider the four most common types of HTTP requests:
\begin{enumerate} 
\item \textbf{GET:} Used to retrieve a resource without making modifications. 
\item \textbf{PUT:} Used to replace a resource’s content with the content of the request. 
\item \textbf{POST:} Used to send data to a resource, triggering processing on the server. 
\item \textbf{DELETE:} Used to remove a resource. 
\end{enumerate}
To assign Java methods to an appropriate HTTP request type, we employ a "Zero-Shot Classification" model. This model predicts the class of an input text without requiring specific training examples for each class. By defining the four classes (''GET," ''PUT," ''POST," ''DELETE") and using information about the method as input text, the model identifies the associated HTTP request type.

Fig.~\ref{fig:restify:querry_classifier} illustrates this process with the method \texttt{java.lang.String getCity()}. The input to the model includes the method's return type, name, and parameters, providing enough context. The model outputs a probability distribution for the four classes. For \texttt{getCity}, the model estimates a 96\% likelihood of belonging to the ''GET" class.

The use of an AI model, rather than direct name-based analysis, is critical due to the wide variety of naming conventions for Java methods. It is impractical to create universal rules for method name classification. For example, while it may be clear that \texttt{getCity} should map to a ''GET" request, another developer might name the same method \texttt{retrieveCity}. Our tool must consistently classify HTTP request types regardless of naming conventions.

\begin{figure}[h]
    \centering
    \includegraphics[width=0.5\textwidth]{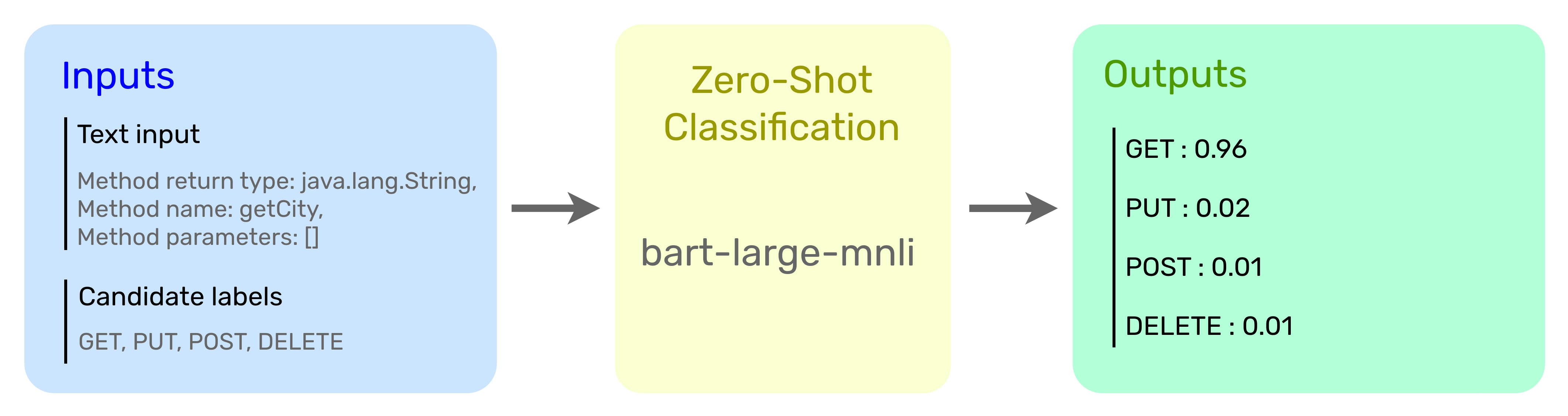}
    \caption{Representation of the HTTP request classification process.}
    \label{fig:restify:querry_classifier}
\end{figure}

The next step in this phase pipeline is the URI creation. A ''Uniform Resource Identifier" (URI) is a string of characters that identifies a resource in a REST API \cite{REST-API-Design-Rulebook}. Each segment of a URI, separated by slashes (/), represents a data hierarchy. In this process, we aim to identify a URI for each method, which will act as the access point for API requests. To achieve this, we utilize several algorithms and machine learning models to generate a URI for each method.

First, we define a data structure in the form of a search tree to store the generated API. Each element in the tree corresponds to a segment of the URI, and its child nodes represent subsequent segments. Each node includes attributes such as the segment name and any associated HTTP requests.
The URI creation process is illustrated in Fig.~\ref{fig:restify:creation_uri_steps}, using methods from a cluster identified in the Spring PetClinic application as an example.

\begin{figure}[h]
    \centering
    \includegraphics[width=0.5\textwidth]{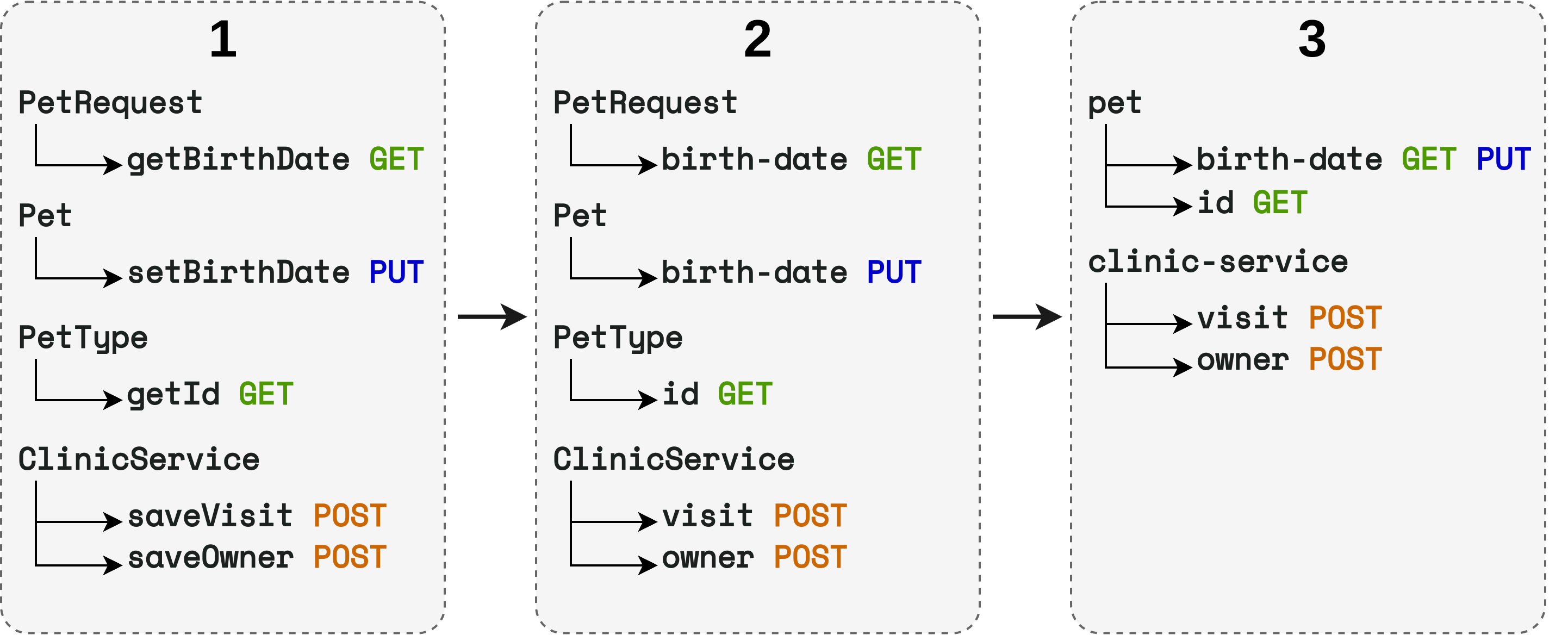}
    \caption{Representation of the URI creation process.}
    \label{fig:restify:creation_uri_steps}
\end{figure}

\begin{enumerate} 
\item \textbf{Tree Initialization:} The tree is initialized with the cluster name as the root node. For each method, a node named after the method's class is created, containing a child node named after the method itself. This corresponds to the first step in Fig.~\ref{fig:restify:creation_uri_steps}. At this stage, the URI format is: \texttt{http://api/cluster/Class/method}. The HTTP requests identified in the previous step are associated with the method nodes.
\item \textbf{Processing Method Name Segments:} URIs must comply with naming conventions \cite{REST-API-Design-Rulebook}. Specifically, URI segments must be lowercase, separated by hyphens for multi-word names, and must not include action verbs. To meet these requirements, action verbs are removed from method name segments, and the results are reformatted to comply with naming conventions. This corresponds to the second step in Fig.~\ref{fig:restify:creation_uri_steps}. Methods are first split into words based on camelCase notation. Then, a "Part-of-Speech tagging" algorithm\footnote{\url{https://www.nltk.org/api/nltk.tag.pos_tag.html}} is applied to identify grammatical roles (e.g., verbs, nouns, adjectives). Verbs are removed when the method name contains more than one word. Finally, the resulting segments are converted to lowercase and formatted with hyphens.

\item \textbf{Processing Class Name Segments:} Several methods may share a common conceptual grouping under a single URI segment associated with their class names. For instance, in the second step of Fig. \ref{fig:restify:creation_uri_steps}, the classes \texttt{PetRequest}, \texttt{Pet}, and \texttt{PetType} all represent the same concept. In a REST API, these methods should be grouped under a single access path. To achieve this, a semantic analysis is performed on URI segments associated with class names. Using the sentence embedding model (SBERT) applied during multi-objective optimization, class names with high semantic similarity are grouped together. A second semantic analysis identifies the most representative word for the grouped classes, which is used as the new URI segment name. The results are formatted to adhere to URI naming conventions. This corresponds to the third step in Fig.~\ref{fig:restify:creation_uri_steps}.
\end{enumerate}


\section{Case Study}
\label{Validation}
The proposed approach is evaluated using the open-source Spring PetClinic AngularJS\footnote{\url{https://github.com/spring-petclinic/spring-petclinic-angularjs}}, a monolithic Java application that includes a server component developed with Java Spring. Additionally, we utilize Spring PetClinic Microservices\footnote{\url{https://github.com/spring-petclinic/spring-petclinic-microservices}}, a version of Spring PetClinic AngularJS migrated to a microservices architecture, consisting of seven microservices. This setup allows us to compare the services identified by our approach on the monolithic application with the microservices in the migrated version, providing a reference for assessing the quality of our solution.

\subsection{First Analysis: Objective Functions and Number of Clusters}
In the first analysis, we examined how well the objectives are achieved for different numbers of clusters. The aim is to demonstrate the importance of setting this parameter to obtain meaningful microservices. Table~\ref{table:optimisation} presents the values of the objective functions for near-optimal solutions obtained after 100 generations, considering 3, 5, 7, and 10 microservices.
\begin{table*}[h]
\centering
    \begin{tabular}{|l|c|c|c|}

  \hline
    \textbf{Number of identified clusters} & \textbf{Coupling} & \textbf{Cohesion} & \textbf{Semantic similarity} \\
   \hline
    3 & $\approx 0.121$ & $\approx 0.667$ & $\approx 0.643$ \\
      \hline
    5 & $\approx 0.209$ & $\approx 0.642$ & $\approx 0.725$ \\
      \hline
   7 & $\approx 0.176$ & $\approx 0.652$ & $\approx 0.757$ \\
      \hline
 10 & $\approx 0.286$ & $\approx 0.542$ & $\approx 0.786$ \\
      \hline
    \end{tabular}
    \caption{Objective function values for varying number of clusters.}
    \label{table:optimisation}
\end{table*}

We observed that the semantic similarity of the solutions increases with the number of microservices. However, coupling and cohesion vary, with the worst values occurring for the largest number of microservices. The best trade-off is achieved with 7 microservices, which matches the number in the reference solution. In the remainder of this section, we focus on the results for this configuration.

Figure~\ref{fig:metrics_evolution} shows the evolution of the three objective functions during the search process for one execution targeting 7 microservices. Initially, the random solutions exhibit the worst values. These values steadily improve over iterations, stabilizing after the 60th generation.

\begin{figure}[h] 
\centering 
\includegraphics[width=0.5\textwidth]{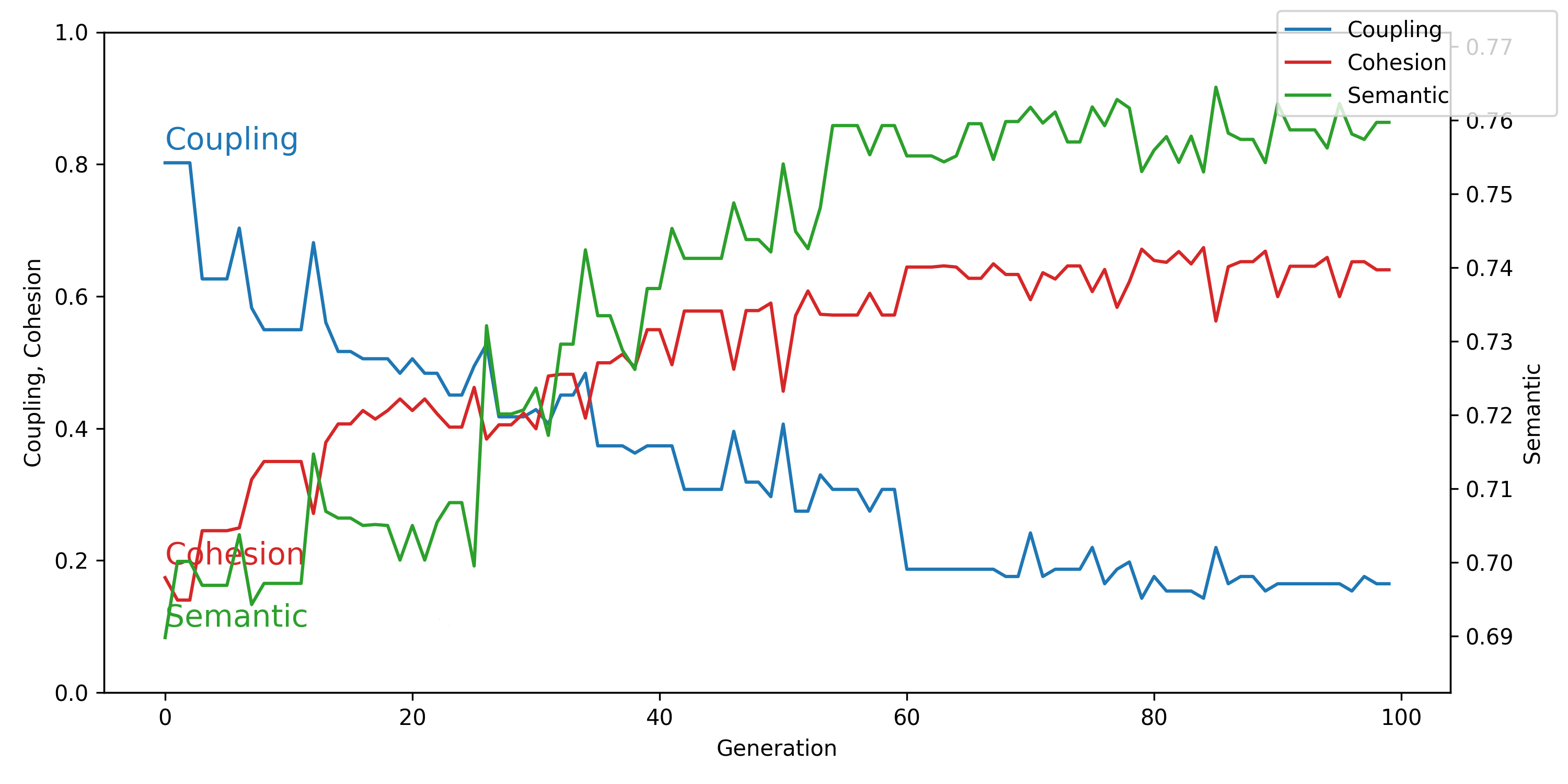}
 \caption{Evolution of objectives through iterations.}
 \label{fig:metrics_evolution}
 \end{figure}
\subsection{Second Analysis: Comparing Clusters to Reference Microervices}
In the second analysis, we compared the clusters identified by our approach to the microservices in the reference version. First, we analyzed the methods present in both applications. While the microservices version retains all functionalities of the monolithic version, the two implementations differ in their structure, resulting in variations in method composition. 

Table~\ref{table:distribution_methods} shows the distribution of methods between the two versions, highlighting that only 22 of the 57 methods from the monolithic application are also found in the microservices version. Consequently, comparing clusters to microservices based solely on method names is not feasible.
To address this limitation, we adopted a semantic comparison approach, as illustrated in Fig.~\ref{fig:validation:validation}. We constructed a similarity matrix by computing the embeddings of the clusters and the microservices in the reference version, using the same embedding model applied earlier for methods. Each cluster and microservice embedding was derived from the combined embeddings of its methods. Clusters were then assigned to microservices based on the highest similarity scores.

\begin{table}[h]
    \centering
    \begin{tabular}{|l|c|}
        \hline
        \textbf{Methods common to both applications} & 22\\ \hline
        \textbf{Methods unique to monolithic application} & 35\\ \hline
        \textbf{Methods unique to the microservices application} & 19\\ \hline
    \end{tabular}
    \caption{Distribution of methods between the monolithic and microservices versions of Spring PetClinic.}
    \label{table:distribution_methods}
\end{table}

The results, shown in Table~\ref{table:validation}, reveal high to very high similarities between the identified clusters and the reference microservices, with values ranging from $0.67$ to $0.9$. Moreover, the identified REST APIs closely align with the microservice interfaces in the reference application.
\begin{figure}[h]
    \centering
    \includegraphics[width=0.5\textwidth]{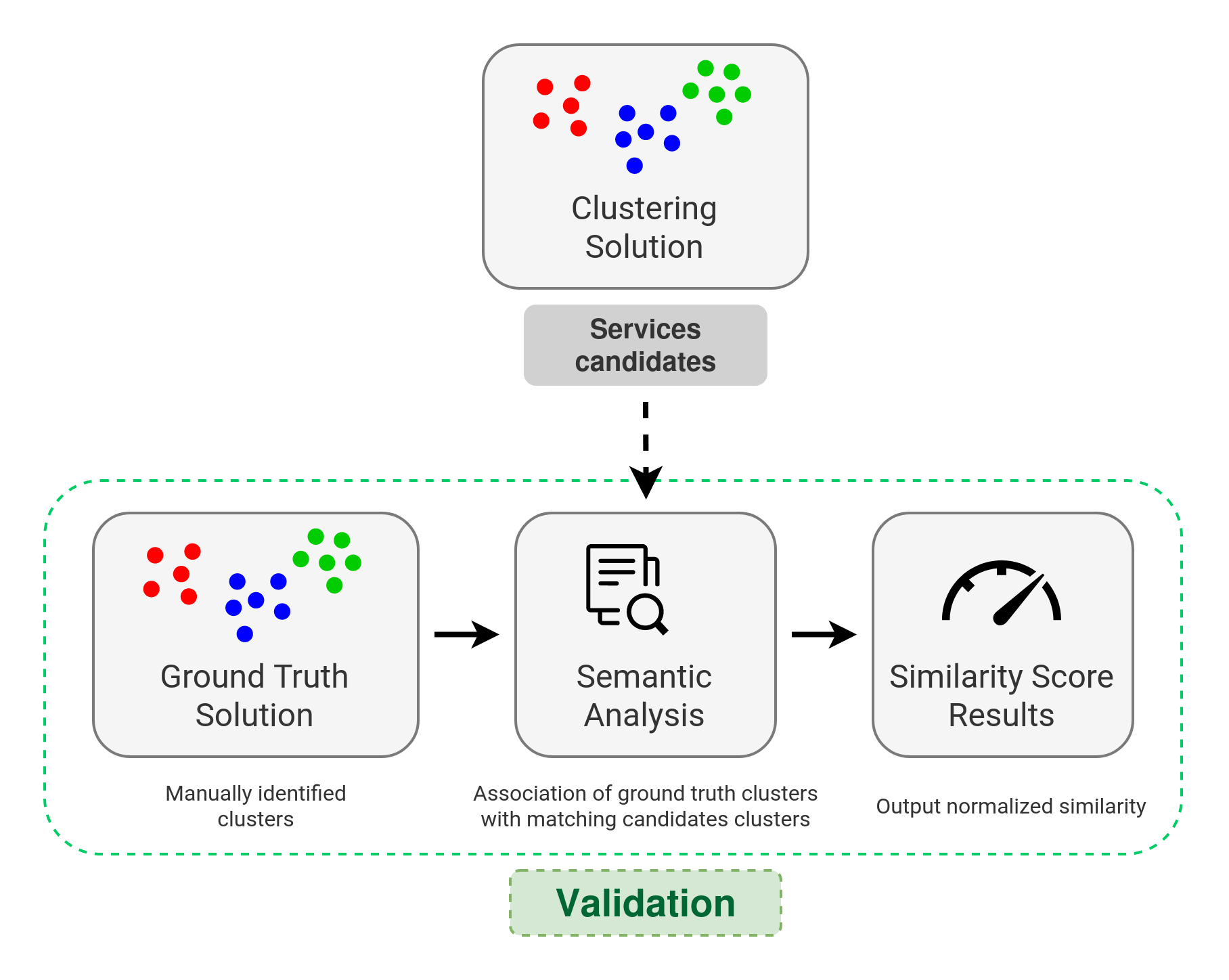}
    \caption{Evaluate strategy for microservice identification.}
    \label{fig:validation:validation}
\end{figure}

\begin{table*}
    \centering
    \begin{tabular}{|l|c|c|c|}    
    \toprule
    \textbf{Identified Clusters} & \centering \textbf{Reference Microservices} & \textbf{Cosine Similarity} \\
\midrule
    Cluster 1 & spring-petclinic-customers-service & $\approx 0.907$ \\
    Cluster 2 & spring-petclinic-config-server & $\approx 0.685$ \\
    Cluster 3 & spring-petclinic-discovery-server & $\approx 0.671$ \\
    Cluster 4 & spring-petclinic-admin-server & $\approx 0.689$ \\
    Cluster 5 & spring-petclinic-vets-service & $\approx 0.786$ \\
    Cluster 6 & spring-petclinic-api-gateway & $\approx 0.866$ \\
    Cluster 7 & spring-petclinic-visits-service & $\approx 0.776$ \\
    \bottomrule
    \end{tabular}
\caption{Similarity between the identified clusters and the reference microservices.}
    \label{table:validation}
\end{table*}

\section{Related work}
\label{RelatedWorks}

Our work aligns with recent efforts to migrate legacy monolithic systems to microservices using multi-objective approaches and machine learning techniques. This section reviews related works selected for their methodological relevance, significance of proposed results, and recency. Service identification methods, central to our research, commonly rely on dependency analysis~\cite{kalia2021mono2micro}, genetic algorithms~\cite{trabelsi2022legacy, MSExtractor}, or machine learning~\cite{nitin2022cargo, MAGNET_ICSA}.

Mono2Micro, a tool developed by IBM, facilitates the migration of monolithic applications to microservices~\cite{kalia2021mono2micro}. It employs a technique based on business use cases and runtime call relationships to generate class-level partitions that are functionally cohesive. Evaluated against four different approaches on various applications, Mono2Micro achieves superior modularity and independence measures, demonstrating its effectiveness.

CARGO, an AI-assisted graph partitioning tool, improves the quality of partitions for microservices migration~\cite{NSGA-III}. Developed as part of IBM’s Project \href{https://github.com/IBM/codenet-minerva}{Minerva}, it employs three main steps: constructing a context-sensitive system dependency graph (SDG), extracting subgraphs to create context-sensitive snapshots, and grouping system components into services. CARGO has been evaluated on five JEE applications, outperforming four state-of-the-art approaches in partition quality.

MSExtractor approaches service identification as a combinatorial optimization problem~\cite{MSExtractor}. Using NSGA-II~\cite{NSGA-II}, it assigns classes to services with the dual objectives of minimizing coupling and maximizing cohesion. Empirical evaluations highlight MSExtractor’s effectiveness, particularly for large-scale systems~\cite{Andritsos2005Information-theoretic, Mazlami2017ExtractionOM, 8456351}. A subsequent enhancement~\cite{ImprovingMSExtractor} replaces NSGA-II with IBEA~\cite{zitzler2004indicator}, introducing an additional objective to optimize service granularity. The updated version demonstrates improved efficiency and produces decompositions with better coupling and cohesion metrics.

Dehghani et al.~\cite{dehghani2022facilitating} propose a reinforcement learning (RL) framework for microservices decomposition. Using the \textsc{ServiceCutter} tool, services are identified and further refined by an RL algorithm that maps methods to microservices. While innovative, the complexity and computational demands of RL models present practical challenges for large-scale systems, requiring significant expertise and resources.

MicroMiner emphasizes the semantic relationships between application elements for microservice identification~\cite{microminer}. Combining static and semantic analysis with machine learning, it evaluates precision and recall on four monolithic systems. Its successor, MAGNET, employs graph neural networks (GNNs)~\cite{MAGNET_ICSA}, integrating static and semantic data into a graph of method calls. MAGNET focuses on service cohesion and low coupling, outperforming MicroMiner in precision and recall across four open-source systems.

Similarly to our work, a microservice identification approach called toMicroservices was proposed, utilizing NSGA-III with five distinct objectives~\cite{assunccao2022analysis}. However, none of these objectives take into account the semantic relationships between the elements of the legacy application, a key aspect addressed in our approach. Building on this work, the same team later explored the perceived usefulness of the identified microservices and found, among other insights, that participants were hesitant to adopt automatically generated microservices~\cite{Carvalho2024}. While this is an interesting finding, its generalizability is limited, as it is closely tied to the specific identification methodology employed by toMicroservices.

In summary, Mono2Micro~\cite{kalia2021mono2micro}, CARGO~\cite{NSGA-III}, and MSExtractor~\cite{ImprovingMSExtractor} operate at the class level, assigning entire classes to single microservices. However, we argue that method-level granularity offers finer decomposition, as methods within the same class may belong to different services, and methods from different classes may contribute to the same service. Class-level partitioning can thus overlook these relationships, limiting the flexibility and accuracy of microservice identification.
Dehghani et al.~\cite{dehghani2022facilitating} present a promising RL-based approach, but the computational complexity of RL models can hinder their scalability and practical application in large systems.
MAGNET aligns closely with our approach in its focus on method-level granularity and semantic integration. However, while MAGNET leverages GNNs, our methodology incorporates a genetic algorithm (NSGA-III), providing a transparent and interpretable optimization process. This transparency allows us to observe the evolution of metrics and solutions, which is less evident in self-learning models like GNNs.

\section{Discussion}
\label{Discussion}
By exposing identified services as microservices without requiring a complete application migration, our approach enables interaction with each service while providing an intermediate step in the migration process. This allows the identified services to remain operational and accessible.

Unlike Mono2Micro, which partitions applications at the class level, our approach focuses on method-level partitioning in monolithic applications. This finer granularity is a key contribution, allowing more precise service identification by grouping methods from different classes into cohesive services when appropriate~\cite{abgaz2023decomposition}. To achieve this, we employ the NSGA-III multi-objective genetic algorithm, which optimizes service partitioning by simultaneously considering several objectives. NSGA-III enables a balanced trade-off among competing goals, ensuring the discovery of high-quality clustering solutions across a diverse solution space~\cite{NSGA-III-implémentation}.

Our contributions include: 
\begin{itemize} 
\item A method-level granularity for service decomposition, which outperforms class-level approaches by capturing finer-grained functional relationships between methods. This ensures that methods contributing to the same business logic can be grouped, even if they belong to different classes. 
\item Integration of a semantic analysis model for method names to enhance service identification. By leveraging semantic similarity, our approach captures the business logic behind methods~\cite{gouigoux2019functionalfirst}, providing critical insights into how the application operates and ensuring meaningful decomposition. 
\end{itemize}

Our approach diverges from existing methods such as MSExtractor, which favors the IBEA algorithm for optimization, and Mono2Micro, which partitions at the class level. While these methods have demonstrated certain advantages, NSGA-III offers greater efficiency and flexibility, particularly for problems involving three or more conflicting objectives~\cite{NSGA-III-comparaison}. Compared to graph neural networks used in approaches like MAGNET, NSGA-III provides transparency in how solutions evolve and metrics improve during execution, making it easier to interpret results.
By combining semantic analysis and the powerful optimization capabilities of NSGA-III, our approach advances the state of the art in microservice identification, offering a more precise and interpretable decomposition strategy for monolithic applications.

\section{Conclusion and Future work}
\label{Conclusion}
In this paper, we proposed a novel approach to migrating legacy monolithic systems to microservices by identifying and exposing services through REST APIs, without requiring a complete application migration. By focusing on method-level granularity and leveraging multi-objective optimization with the NSGA-III genetic algorithm, our approach achieves more accurate service identification compared to traditional class-level partitioning methods. We also integrated semantic analysis to better capture the business logic, enhancing the decomposition process and improving the quality of the resulting microservices.

The evaluation of our approach on the Spring PetClinic application demonstrated its effectiveness in identifying meaningful services that closely resemble the structure and functionality of microservices in the reference version. The results showed promising semantic similarity between the identified clusters and the actual microservices, highlighting the potential of our method to facilitate gradual migrations and serve as a viable intermediate step toward full microservice adoption.

While the results are encouraging, there are several avenues for future work that could further refine and expand the applicability of our approach. Although our method focuses on method-level decomposition based on method calls, other fine-grained analyses could be incorporated, such as integrating data access patterns or dynamic analysis to uncover usage patterns.
Additionally, the semantic analysis can be enhanced by incorporating more advanced Natural Language Processing (NLP) techniques or by using pre-trained models specifically tailored for software engineering tasks. Exploring domain-specific semantic models could also improve the identification of business logic and service boundaries.

Although NSGA-III offers significant advantages in multi-objective optimization, integrating other optimization techniques—such as deep reinforcement learning or graph-based optimization—could further refine the clustering process, particularly in dynamic environments where system requirements evolve over time.
Incorporating continuous learning represents another promising direction for future research. Future iterations of the approach could benefit from feedback loops, where the system learns from the successes or failures of prior service compositions. This could involve the use of reinforcement learning or active learning techniques to optimize service identification and API exposure as the system evolves.

Furthermore, expanding our approach to automatically generate API documentation or refine the structure of the REST API based on the identified services would provide a more comprehensive solution for microservice exposure.

Finally, extending the evaluation to more complex, real-world legacy systems would provide deeper insights into the scalability and robustness of our approach in diverse environments, particularly for highly complex monolithic applications.

\section*{Acknowledgment}
This work is partially supported by the Natural Sciences and Engineering Research Council of Canada under Grant No. RGPIN-2019-07168 and the IUT de Nantes (Nantes Université).

\bibliographystyle{plain}
\balance
\bibliography{biblio}

\end{document}